\documentclass[11pt]{article}
\usepackage[utf8]{inputenc}
\pdfoutput=1
\usepackage{setspace}
\usepackage{palatino}
\usepackage{graphicx}
\usepackage{adjustbox}
\usepackage{subfigure}
\usepackage{float}
\usepackage{titling} 
\usepackage{multirow}
\usepackage{lscape}
\usepackage{amsmath}
\usepackage{amssymb}
\usepackage{mathtools}
\usepackage{booktabs}
\usepackage{rotating}
\usepackage{amsmath}
\usepackage{pdfpages}
\usepackage{float}
\usepackage{booktabs,caption}
\usepackage[flushleft]{threeparttable}
\usepackage[utf8]{inputenc}
\usepackage{enumitem}
\usepackage{placeins}
\usepackage{breqn}
\usepackage{enumitem}
\usepackage[a4paper, total={6in, 9.5in}]{geometry}
\fontfamily{ppl}\selectfont 
\usepackage[american]{babel}
\usepackage{bm}
\usepackage{caption}
\usepackage{array}
\usepackage{booktabs}
\usepackage{dcolumn}
\usepackage{graphicx}
\newcolumntype{d}{D{.}{.}{2.5}}           
\usepackage[group-separator={,}]{siunitx}
\usepackage[normalem]{ulem}
\usepackage[textsize=small]{todonotes}
\usepackage{makecell}

\newcommand{\q}[1]{'#1'}

\usepackage[natbibapa]{apacite}
\PassOptionsToPackage{hyperindex,breaklinks}{hyperref}
\usepackage{hyperref}
\usepackage{xcolor}
\definecolor{darkblue}{rgb}{0, 0, 0.5}
\hypersetup{colorlinks=true,citecolor=darkblue, linkcolor=darkblue, urlcolor=darkblue}
\usepackage{lineno}

\begin{document}

\title{Nowcasting in triple-system estimation}

\author{Daan B. Zult\\
   {\small\raggedright Statistics Netherlands}\\
   \href{mailto:db.zult@cbs.nl}{\texttt{db.zult@cbs.nl}}
\and Peter G. M. van der Heijden\\
    {\small\raggedright Utrecht University and University of Southampton}\\
    \href{mailto:P.G.M.vanderHeijden@uu.nl}{\texttt{P.G.M.vanderHeijden@uu.nl}}
\and Bart F. M. Bakker\\
    {\small\raggedright Statistics Netherlands and VU University Amsterdam}\\
    \href{mailto:bfm.bakker@cbs.nl}{\texttt{bfm.bakker@cbs.nl}}
}

\date{}
\date{\vspace{-5ex}}

\renewenvironment{abstract}
 {\par\noindent\textbf{\abstractname: }\ \ignorespaces}
 {\par\medskip}

{\setstretch{.8}
\maketitle
\begin{abstract}
Multiple systems estimation uses samples that each cover part of a population to obtain a total population size estimate. Ideally, all the available samples are used, but if some samples are available (much) later, one may use only the samples that are available early. Under some regularity conditions, including sample independence, two samples is enough to obtain an asymptotically unbiased population size estimate. However, the assumption of sample independence may be unrealistic, especially when samples are derived from administrative sources. The sample independence assumption can be relaxed when three or more samples are used, which is therefore generally recommended. This may be a problem if the third sample is available much later than the first two samples. Therefore, in this paper we propose a new approach that deals with this issue by utilising older samples, using the so-called expectation maximisation algorithm. This leads to a population size nowcast estimate that is asymptotically unbiased under more relaxed assumptions than the estimate based on two samples. The resulting nowcasting model is applied to the problem of estimating the number of homeless people in The Netherlands, which leads to reasonably accurate nowcast estimates.

\end{abstract}
}

\section{Introduction}  \label{sec:intro}
A well-known problem in the production of statistics is that data may become available gradually, while it may be desirable to produce this statistic before all these data are available. In such cases, it is common practice to produce a preliminary statistic that can also be referred to as a nowcast, based on some part of the data that is available early, and update this statistic when all data is available. Discussions on this topic usually evolve around correcting for response bias that may occur when the speed of response is related to the statistic itself. For example, when companies with a quickly growing turnover also respond quickly, a nowcast on turnover growth might be biased upwards if this relation is ignored.

A statistic for which such a nowcasting method is not available is a population size estimate that is based on samples that each partly observe a population, and some of these (complete) samples are available with delay. This may occur when, for example, samples are registers or surveys that are maintained or collected periodically throughout a certain period, where the statistician has little control over the data collection process. Then, some samples might be available early and others later. In such cases it is common practice to simply wait until all samples have become available before estimation is performed. This raises the question whether and under what conditions it is possible to produce a preliminary population size estimate based on the set of samples that are available earlier. The simplest case is when one sample becomes available early and a second sample becomes available later. However, when these two samples are not independent, which is often the case when samples are derived from administrative sources, more than two samples are required to obtain an asymptotically unbiased population size estimate. This brings us to the the more complex case, and the main topic of this paper, what if three samples become available sequentially and a statistic needs to be produced after the first or second sample becomes available?

The models that are involved in the estimation of the size of a partly observed population are known under different names such as capture-recapture, mark and recapture or multiple systems estimation (MSE). When the number of samples is two, MSE is usually referred to as dual-system estimation (DSE) and when the number of samples is three, triple-system estimation (TSE). The most basic DSE model was proposed by \cite{Petersen1896}, and later by \cite{Lincoln1930}. Under a set of assumptions discussed by \cite{Wolter1986}, a DSE estimator provides an asymptotically unbiased population size estimate. When samples are derived from administrative sources, two of these assumptions become problematic. The first is the DSE assumption of homogeneous inclusion probabilities in at least one sample \cite[see e.g.][]{Seber1982,vdHeijden2012}, which can be somewhat relaxed when individual background variables/covariates (e.g. gender, age, etc.), that are related to inclusions probabilities, can be included in the the model. The second problematic DSE assumption is the assumption of sample independence, which can be relaxed when three or more samples are included in the model (see Section \ref{sec:DSE} and \ref{sec:TSE} for further details). This implies that if samples are derived from administrative sources, TSE is generally recommended over DSE. 

The basics of TSE are discussed in \cite{Fienberg1972}, and it has a long history in informing public policy \cite[see e.g.][for an extensive discussion]{Bird2018}. For further technical discussions and applications of TSE with administrative data, see e.g. \cite{Baffour2013,Gerritse2016,Baffour2021}, who apply TSE to estimate the size of an unobserved part of a population in case of census or register data, or see \cite{Heijden2020} for an MSE application with four samples, including the population census, to estimate the size of the Māori population in New Zealand.

The case considered in this paper is that of a contingency table based on three samples for the period, and a contingency table based on one or two samples for the current period is available. The goal is to obtain a maximum likelihood (ML) population size estimate for the current period. The absence of a second and third, or only a third sample, for the current period could be considered a missing or incomplete data problem. A standard method to deal with incomplete data is the expectation–maximization (EM) algorithm \citep{Dempster1977}. The EM algorithm method allows for statistical inference from incomplete data with ML. In this paper we will discuss under which conditions the EM algorithm can be combined with DSE and TSE to obtain an asymptotically unbiased population size nowcast (NC) estimate. This approach of combining the EM algorithm with MSE models based on incomplete data is not new. For example, \cite{Zwane2004} consider the case that some samples may contain different but overlapping populations, \cite{Zwane2007} consider the case where some covariates are missing in some samples, and \cite{Heijden2020} who applies MSE and uses the EM algorith to complete some missing data. New in this study is that the method is applied to obtain nowcasts for which both observations and estimates based on fully observed MSE data become available later. This allows us to compare the nowcasting model estimates with actual observations and estimates that are based on fully observed MSE data in a practical example.

Next, Section \ref{sec:notation} discusses some basics of the DSE and TSE model, and how data for two periods can be combined in one framework. Next, Section \ref{sec:EM} discusses how the EM algorithm can be used to obtain ML estimates within this framework with incomplete data. The combination of DSE and TSE models and the EM algorithm, leads to our proposed MSE nowcasting model. Finally, in Section \ref{sec:homeless} the nowcasting model is applied to obtain NC estimates for the number of homeless people in The Netherlands. These NC estimates are compared with alternative estimates such as standard DSE estimates.

\section{Theory and notation} \label{sec:notation}

This section discusses DSE and TSE notation and theory, and shows how DSE and TSE models can be combined over two periods. 

\subsection{Dual-system estimation} \label{sec:DSE}
Imagine a population with size $N$ and a set of two samples $A$ and $B$ that each cover part of this population. The goal is to use these samples to obtain a population size estimate denoted as $\hat{N}$. When each unit in each sample can be uniquely identified, then for each unit an inclusion pattern $ab$ can be constructed, with $a,b \in \{ 0,1 \}$, where $a=0$ stands for \q{\textit{not included in sample $A$}} and $a=1$ for \q{\textit{included in sample $A$}}, and the same with $b$ for sample $B$. The units of each inclusion pattern can be counted and denoted as $n_{ab}$, except when the inclusion pattern is $00$, because these units are unobserved. The sum of all observed units is denoted as $n$ and so $n = n_{11} + n_{10} + n_{01}$. Finally, when we sum over $a$ or $b$, we replace that subscript by a \q{\texttt{+}}. Thus, for example, $n_{1\texttt{+}} = n_{10} + n_{11}$ is equal to the size of source $A$. It is assumed that $n_{ab}$ is a realisation of a random variable with expectation $m_{ab}$ and the aim of DSE is to obtain $\hat{m}_{ab}$, an estimate of this expectation. 

Under a set of assumptions discussed by for example,  \cite{Wolter1986}, the observed counts $n_{11}$, $n_{10}$ and $n_{01}$ can be used to estimate $N$. These assumptions can be summarised as:
\begin{enumerate}
\item The sampling population is equal for samples $A$ and $B$.
\item Records that correspond to the same unit in samples $A$ and $B$ can be perfectly linked.
\item Inclusion probabilities are homogeneous in samples $A$ or $B$ \citep[see e.g.][]{Seber1982}.
\item Samples $A$ and $B$ are independent. 
\end{enumerate}
Under assumption (1-4), an asymptotically unbiased DSE-estimator for $m_{00}$ can be written as
\begin{align} \label{eq:DSE}
\hat{m}^{\text{DSE}}_{00} = & \frac{n_{10}n_{01}}{n_{11}}, &&
\end{align}
and consequently for $N$ as $\hat{N}^{\text{DSE}} = n + \hat{m}^{\text{DSE}}_{00} = \frac{n_{1\texttt{+}}n_{\texttt{+}1}}{n_{11}}$.

\cite{Fienberg1972} showed that the DSE estimator can also be derived from a log-linear model for $m_{ab}$, and for our purpose it is important to show how this relates to the independence assumption 4. A log-linear model for $m_{ab}$ can be written as
\begin{align} \label{eq:LL2}
\log m_{ab} = \lambda + \lambda_{a}^{A} + \lambda_{b}^{B} + \lambda_{ab}^{AB}, &&
\end{align}
with $\lambda$ an intercept term, $\lambda_{a}^{A}$ and $\lambda_{b}^{B}$ are the respective inclusion parameters for samples $A$ and $B$ that are identified by setting $\lambda_{0}^{A} = \lambda_{0}^{B} = 0$ and $\lambda_{ab}^{AB}$ is a parameter for the interaction between samples $A$ and $B$. Because $m_{00}$ is unobserved and the independence assumption 4 implies that $\lambda_{ab}^{AB}=0$, in practice the expectations $m_{ab}$ are replaced with the observed counts/their ML estimates $n_{ab}$ after which Eq. (\ref{eq:LL2}) represents three equations and three unknowns that lead to the DSE-estimator in Eq. (\ref{eq:DSE}). This also shows that if $\lambda_{ab}^{AB} \neq 0$, then $\hat{m}_{00}^{\text{DSE}}$ is a biased estimate for $m_{00}$. In the next section we will show how TSE may solve this problem of bias due to pairwise dependence of samples.

\subsection{Triple-system estimation} \label{sec:TSE}

When instead of being observed by two samples, a population is partly observed by three samples $A$, $B$ and $C$, each unit has an inclusion pattern that, instead of $ab$, can be written as $abc$, where $c$ is defined in the same way as $a$ and $b$. This means that instead of the four inclusion patterns in DSE there are now eight TSE inclusion patterns $000$, $100$, $010$, $001$, $110$, $101$, $011$ and $111$, and Eq. (\ref{eq:LL2}) can be extended towards
\begin{align} \label{eq:LL3}
\log m_{abc} = \mu + \mu_{a}^{A} + \mu_{b}^{B} + \mu_{c}^{C} + \mu_{ab}^{AB} + \mu_{ac}^{AC} + \mu_{bc}^{BC} + \mu_{abc}^{ABC}. && 
\end{align}

Eq. (\ref{eq:LL3}) is the three-sample representation of Eq. (\ref{eq:LL2}), with $\mu$ the intercept, $\mu_{a}^{A}$, $\mu_{b}^{B}$ and $\mu_{c}^{C}$ the inclusion parameters, $\mu_{ab}^{AB}$, $\mu_{ac}^{AC}$ and $\mu_{bc}^{BC}$ the pairwise interaction parameters and $\mu_{abc}^{ABC}$ a three-way interaction parameter. It constitutes a system of eight linear equations and eight unknowns, but because $n_{000}$ is unobserved, it cannot be solved when $m_{abc}$ are replaced by the observed counts/their ML estimates $n_{abc}$. Therefore it is usually assumed that $\mu_{abc}^{ABC} = 0$, which is similar to DSE assumption 4 but more realistic. This assumption gives the so-called saturated TSE model
\begin{align} \label{eq:LL3_SAT}
\text{saturated: } \log m_{abc} = \mu + \mu_{a}^{A} + \mu_{b}^{B} + \mu_{c}^{C} + \mu_{ab}^{AB} + \mu_{ac}^{AC} + \mu_{bc}^{BC}, && 
\end{align}
\noindent that in contrast to DSE, also contains pairwise interaction parameters $\mu_{ab}^{AB}$, $\mu_{ac}^{AC}$ and $\mu_{bc}^{BC}$. 

This model can be further restricted by setting one or more pairwise interaction terms to zero, which gives seven additional models, i.e.:
\begin{align}
\text{two-pair dependence (I): } & \log m_{abc} = \mu + \mu_{a}^{A} + \mu_{b}^{B} + \mu_{c}^{C} + \mu_{ac}^{AC} + \mu_{bc}^{BC},  && \label{eq:LL3_2PD_I} \\
\text{two-pair dependence (II): } & \log m_{abc} = \mu + \mu_{a}^{A} + \mu_{b}^{B} + \mu_{c}^{C} + \mu_{ab}^{AB} + \mu_{bc}^{BC}, \label{eq:LL3_2PD_II} \\
\text{two-pair dependence (III): } & \log m_{abc} = \mu + \mu_{a}^{A} + \mu_{b}^{B} + \mu_{c}^{C} + \mu_{ab}^{AB} + \mu_{bc}^{AC}, \label{eq:LL3_2PD_III} \\
\text{one-pair dependence (I): } & \log m_{abc} = \mu + \mu_{a}^{A} + \mu_{b}^{B} + \mu_{c}^{C} + \mu_{bc}^{BC}, \label{eq:LL3_1PD_I} \\
\text{one-pair dependence (II): } & \log m_{abc} = \mu + \mu_{a}^{A} + \mu_{b}^{B} + \mu_{c}^{C} + \mu_{ac}^{AC}, \label{eq:LL3_1PD_II} \\
\text{one-pair dependence (III): } & \log m_{abc} = \mu + \mu_{a}^{A} + \mu_{b}^{B} + \mu_{c}^{C} + \mu_{ab}^{AB}, \label{eq:LL3_1PD_III} \\
\text{independence: } & \log m_{abc} = \mu + \mu_{a}^{A} + \mu_{b}^{B} + \mu_{c}^{C}. \label{eq:LL3_IND}
\end{align}
Models with more than three samples can be developed along the same lines. For an extensive discussion of the models in Eq. (\ref{eq:LL3_SAT}-\ref{eq:LL3_IND}), we refer to \cite{Fienberg1972} and \cite{BishopFienberg1975}, who also discuss derivations of closed form expressions or approximations for $\hat{N}^{\text{TSE}}$ together with asymptotic variances, for each of these models. The distinction between the different restricted models is important when TSE and DSE over two periods are combined. This will be discussed in the next section. 

\subsection{Combining samples over two periods.} \label{sec:combine}

We consider a population with size $N_{t}$ and the samples $A_{t}$, $B_{t}$ and $C_{t}$ that each cover parts of this population for time $t$. Also assume the delivery dates $t=t_{0},t_{1,a},t_{1,b},t_{1,c}$ where at $t_{0}$ the samples $A_{t_{0}}$, $B_{t_{0}}$ and $C_{t_{0}}$ for time $t=t_{0}$ are available and at delivery dates $t_{1,a},t_{1,b}$ and $t_{1,c}$ the samples $A_{t_{1}}$, $B_{t_{1}}$ and $C_{t_{1}}$ for time $t=t_{1}$ become available, one-by-one, in that order. This means that at both $t=t_{0}$ and $t=t_{1,c}$ three samples are available for their corresponding periods $t_{0}$ and $t_{1}$. When we write $abc,t$ as the inclusion pattern for time $t$, a table can be constructed that shows which observed counts are available at which moment, as in Table \ref{tab:combined} below.

\begin{table}[!htb]
    \caption{The observed counts $n_{abc,t_{0}}, n_{abc,t_{1,a}}, n_{abc,t_{1,b}}, n_{abc,t_{1,c}}$ available at the four different moments $t=t_{0}$, $t_{1,a}$, $t_{1,b}$ and $t_{1,c}$.} \label{tab:combined}
        \begin{tabular}{l|l}
                \toprule
                $t$ & The available observed counts $n_{abc,t}$ \\
                \midrule
                $t_{0}$ & $n_{abc,t_{0}}$ \\
                $t_{1,a}$ & $n_{abc,t_{0}}$ and $n_{1\texttt{+}\texttt{+},t_{1}}$ \\
                $t_{1,b}$ & $n_{abc,t_{0}}$ and $n_{ab\texttt{+},t_{1}}$ \\
                $t_{1,c}$ & $n_{abc,t_{0}}$ and $n_{abc,t_{1}}$ \\
                \bottomrule
            \end{tabular}
\end{table}

Table \ref{tab:combined} shows that for $t=t_{0}$ and $t=t_{1,c}$ all observed counts are available for their corresponding periods, and so for each period a TSE-estimate for $m_{000,t}$, as discussed in Section \ref{sec:TSE}, can be estimated. We write their corresponding TSE models as $M_{t_{0}}(\boldsymbol{\mu}_{t_{0}})$ and $M_{t_{1,c}}(\boldsymbol{\mu}_{t_{1,c}}) = M_{t_{1}}(\boldsymbol{\mu}_{t_{1}})$ with $\boldsymbol{\mu}_{t}$ as the vector of $\mu_{t}$-parameters at time $t$. This gives, for example, for the saturated model in Eq. (\ref{eq:LL3_SAT}) at $t=t_{0}$, the set \\ $\boldsymbol{\mu_{t_{0}}} = (\mu_{t_{0}}, \mu_{a,t_{0}}^{A}, \mu_{b,t_{0}}^{B}, \mu_{c,t_{0}}^{C}, \mu_{ab,t_{0}}^{AB}, \mu_{ac,t_{0}}^{AC}, \mu_{bc,t_{0}}^{BC})$.

At $t=t_{1,b}$ and $t=t_{1,a}$ it is not possible to apply TSE for $t=t_{1}$, because at those delivery dates one or two samples are still missing. Table \ref{tab:combined} shows that at those moments only aggregated observed counts are available. Then the question becomes if and under which assumptions, the old samples $A_{t_{0}}$, $B_{t_{0}}$ and $C_{t_{0}}$, together with the aggregated observed counts for $t=t_{1}$, can be used to obtain an asymptotically unbiased estimate for $N_{t_{1}}$. In general, for each observed count that corresponds to a period $t$, one additional parameter for that period can be estimated. This reasoning allows us to construct different MSE models based on the available samples at a given time. 

At $t=t_{1,a}$ the additional observed count $n_{1\texttt{++},t_{1}}$ becomes available, which simply is the total sample size of $A_{t_{1}}$. This can be considered one observed count for time $t=t_{1}$ and therefore allows a model with one additional parameter for time $t=t_{1}$, i.e.
\begin{align}\label{eq:LL3_13}
M_{t_{1,a}}(\boldsymbol{\mu_{t_{1,a}}}) = \log m_{abc,t} = M_{t_{0}}(\boldsymbol{\mu_{t_{1,a}}}) + \mu_{t_{1}}, && 
\end{align}
with $M_{t_{0}}(\boldsymbol{\mu_{t_{0}}})$ as one of the models in Eq.\ (\ref{eq:LL3_SAT} - \ref{eq:LL3_IND}), but with a $t_{0}$ as subscript to the intercept parameter $\mu$. This implies that in $M_{t_{1,a}}(\boldsymbol{\mu_{t_{1,a}}})$, the intercept term may differ for $t=t_{0}$ and $t=t_{1}$, but the inclusion and interaction parameters are equal for both periods. This means that for $m_{000,t_{1}}$, Eq.\ (\ref{eq:LL3_13}) reduces to the expression $m_{000,t_{1}} = \exp \left(\mu_{t_{1}} \right)$. This leads to an asymptotically unbiased estimate for $m_{000,t_{1}}$ and therefore $N_{t_{1}}$ if the estimated inclusion and interaction parameters are independent of $t$.

At $t=t_{1,b}$ the additional sample $B_{t_{1}}$ becomes available and so at $t=t_{1,b}$ two samples are available for time $t=t_{1}$. Table \ref{tab:combined} shows that this means that three observed counts, with inclusion patterns $ab\texttt{+} = 11\texttt{+},10\texttt{+},01\texttt{+}$, are available. This implies that for time $t=t_{1}$ a DSE-estimate can be obtained, but as was discussed in Section \ref{sec:DSE}, this estimate is biased if the independence assumption is violated (i.e. if $\lambda_{ab}^{AB} \neq 0$). Then the question becomes if the presence of the samples $A_{t_{0}}$, $B_{t_{0}}$ and $C_{t_{0}}$ allows for a way in which the independence assumption can be relaxed. Note that due to the three observed counts we can extend $M_{t_{1,a}}(\boldsymbol{\mu_{t_{1,a}}})$ in Eq. (\ref{eq:LL3_13}) with two additional parameters for $t=t_{1}$, i.e.
\begin{align}\label{eq:LL3_23}
M_{t_{1,b}}(\boldsymbol{\mu_{t_{1,b}}}) = \log m_{abc,t} = M_{t_{0}}(\boldsymbol{\mu_{t_{1,b}}}) + \mu_{t_{1}} + \mu_{a,t_{1}}^{A}  + \mu_{b,t_{1}}^{B}. && 
\end{align}
$\boldsymbol{\mu_{t_{1,b}}}$ differs from the $\boldsymbol{\mu_{t_{1,a}}}$ in Eq. (\ref{eq:LL3_13}), because instead of the time independent parameters $\mu_{a}$ and $\mu_{b}$, it contains the time dependent parameters $\mu_{a,t_{0}}$ and $\mu_{b,t_{0}}$, which are just like $\mu_{t_{0}}$ equal to zero for $t=t_{0}$.
This model again gives the same expression $\exp \left(\mu_{t_{1}}\right)$, but the conditions under which the ML-estimator for the parameter $\mu_{t_{1}}$ is an asymptotically unbiased estimator are more relaxed. Note that the parameters in $M_{t_{0}}(\boldsymbol{\mu_{t_{1,b}}})$ that should hold for both periods have reduced with the addition of $\mu_{a,t_{0}}^{A}$ and $\mu_{b,t_{0}}^{B}$, which now, due to the presence of $\mu_{a,t_{1}}^{A}$ and $\mu_{b, t_{1}}^{B}$, correspond exclusively to inclusion probabilities for time $t_{0}$. Therefore, for model $M_{t_{1,b}}(\boldsymbol{\mu_{t_{1,b}}})$ to hold, as compared to model $M_{t_{1,a}}(\boldsymbol{\mu_{t_{1,a}}})$, only the inclusion parameter $\mu_{c}$ and the interaction parameters $\mu_{ab}^{AB}$, $\mu_{ac}^{AC}$ and $\mu_{bc}^{BC}$ should be independent of $t$.

When $m_{abc,t} = m_{ab}$, $\mu_{t_{1}} = \lambda$, $\mu_{a,t_{1}}^{A} = \lambda_{a}^{A}$, $\mu_{b,t_{1}}^{B} = \lambda_{b}^{B}$ and $M_{t_{0}}(\boldsymbol{\mu_{t_{1,b}}}) = \lambda_{ab}^{AB}$, the MSE model in Eq.\ (\ref{eq:LL3_23}) and the DSE model in Eq.\ (\ref{eq:LL2}) are equivalent. This implies that for $M_{t_{1,b}}(\boldsymbol{\mu_{t_{1,b}}})$ the DSE independence assumption 4. can be replaced by the (more relaxed) assumption
\begin{enumerate}
\item[4.] The pairwise dependence parameter $\lambda_{ab}^{AB}$ is independent of $t$.
\end{enumerate}
In other words, the estimate for $\lambda_{ab}^{AB} = \mu_{c} + \mu_{ab}^{AB} + \mu_{ac}^{AC} + \mu_{bc}^{BC}$ for the previous period, can be used as an estimate for the current period, because it is assumed to be stable between both periods.

The estimation of the parameters in the models $M_{t_{1,a}}(\boldsymbol{\mu_{t_{1,a}}})$ and $M_{t_{1,b}}(\boldsymbol{\mu_{t_{1,b}}})$ is less straightforward than the estimation of the parameters in $M_{t_{0}}(\boldsymbol{\mu_{t_{0}}})$ and $M_{t_{1}}(\boldsymbol{\mu_{t_{1}}})$, which can be estimated directly with ML. How to deal with this problem is discussed in the next section.

\section{Combining DSE and TSE with the EM algorithm}\label{sec:EM}

Table \ref{tab:combined} from the previous section poses two statistical estimation problems. On top of the problem of the unobserved counts $n_{000,t_{0}}$ and $n_{000,t_{1}}$, it also poses a so-called mixture model problem \cite[see e.g.][]{Lindsay1995}. This problem implies that for (some) variables only an aggregate over different groups is observed, or one may say that for some groups the data is incomplete. In this case, at $t=t_{1,a}$, there is the aggregated observed count $n_{1\texttt{++},t_{1}}$ and at $t=t_{1,b}$ there are the three aggregated observed counts $\left(n_{11\texttt{+},t_{1}},n_{10\texttt{+},t_{1}},n_{01\texttt{+},t_{1}}\right)$. $n_{1\texttt{++},t_{1}}$ is simply the size of sample $A_{t_{1}}$, and $\left(n_{11\texttt{+},t_{1}},n_{10\texttt{+},t_{1}},n_{01\texttt{+},t_{1}}\right)$ are the aggregated observed counts over sample $C_{t_{1}}$ of the units included in sample $A_{t_{1}}$ and/or $B_{t_{1}}$. A standard method to deal with incomplete data is the EM algorithm. In this case it allows for the estimation of the underlying counts that together add up to the observed aggregated counts, such as the unobserved $n_{111,t_{1}}$ and $n_{110,t_{1}}$ at $t=t_{1,b}$ that add up to the observed $n_{11\texttt{+},t_{1}}$. 

The EM algorithm was introduced by \cite{Dempster1977} as a tool to obtain ML-estimates in case of incomplete data due to unobserved or latent variables. 
In the problem discussed in this paper, the EM algorithm can be applied with model $M_{t_{1,a}}(\boldsymbol{\mu_{t_{1,a}}})$ or $M_{t_{1,b}}(\boldsymbol{\mu_{t_{1,b}}})$ in Eq. (\ref{eq:LL3_13}) and (\ref{eq:LL3_23}). This gives, based on the model for $t_{0}$, an estimated split-up of the available frequencies at $t_{1}$ (e.g. $n_{1\texttt{+}\texttt{+},t_{1,a}} = \hat{n}_{111,t_{1}}+\hat{n}_{110,t_{1}}+\hat{n}_{101,t_{1}}+\hat{n}_{100,t_{1}}$), which we refer to as completed counts. For this case, the outcome of the EM algorithm at $t=t_{t,a}$ and $t=t_{t,b}$ is shown in Table \ref{tab:EM}.

\begin{table}[!htb]
    \caption{Table with completed counts at $t = t_{0}, t_{1,a}, t_{1,b}$.} \label{tab:EM}
        \begin{tabular}{l|l}
                \toprule
                $t$ & The available and completed counts \\
                \midrule
                $t_{0}$ & $n_{abc,t_{0}}$ \\
                $t_{1,a}$ & $n_{abc,t_{0}}$, $\hat{n}_{111,t_{1}}$, $\hat{n}_{110,t_{1}}$, $\hat{n}_{101,t_{1}}$ and $\hat{n}_{100,t_{1}}$ \\
                $t_{1,b}$ & $n_{abc,t_{0}}$, $\hat{n}_{111,t_{1}}$, $\hat{n}_{110,t_{1}}$, $\hat{n}_{101,t_{1}}$, $\hat{n}_{100,t_{1}}$, $\hat{n}_{011,t_{1}}$ and $\hat{n}_{010,t_{1}}$ \\
                \bottomrule
            \end{tabular}
\end{table}

To illustrate how the Expectation step (E-step) of the EM algorithm yields completed data in the columns $\hat{n}_{abc,t_{1,a}}$ and $\hat{n}_{abc,t_{1,b}}$ in Table \ref{tab:EM}, we discuss this for $\hat{n}_{abc,t_{1,b}}$. The EM algorithm allows to split-up $n_{ab\texttt{+},t_{1}}$ into the completed data $\hat{n}_{ab1,t_{1,b}}$ and $\hat{n}_{ab0,t_{1,b}}$ with $\hat{n}_{ab1,t_{1,b}}+\hat{n}_{ab0,t_{1,b}} = n_{ab\texttt{+},t_{1}}$. The EM algorithm starts with an initialisation step that creates an initial set of completed data by, for example, $\hat{n}_{ab1,t_{1,b}}^{(0)} = n_{ab\texttt{+},t_{1}}/2$ and $\hat{n}_{ab0,t_{1,b}}^{(0)} = n_{ab\texttt{+},t_{1}}/2$. Next, in the first maximisation step (M-step) these completed data are assumed regular observations that, together with $n_{abc,t_{0}}$, can be used to estimate the parameters of the model $M_{t_{1,b}}(\boldsymbol{\mu_{t_{1,b}}})$ in Eq. (\ref{eq:LL3_23}), but here it is also possible to replace $M_{t_{0}}(\boldsymbol{\mu_{t_{0}}})$ with a more restricted model. The model resulting from this application of the EM algorithm gives, at iteration $0$, the fitted values $\hat{m}_{abc,t}^{(0)}$. Next, in the first expectation step (E-step) these fitted values are used to split-up $n_{ab\texttt{+},t_{1}}$ (again), but now as $\hat{n}_{ab1,t_{1,b}}^{(1)} = n_{ab\texttt{+},t_{1}} (\hat{m}_{ab1,t_{1,b}}^{(0)}/\hat{m}_{ab\texttt{+},t_{1,b}}^{(0)}$) and $\hat{n}_{ab0,t_{1,b}}^{(1)} = n_{ab\texttt{+},t_{1}} (\hat{m}_{ab0,t_{1,b}}^{(0)} /\hat{m}_{ab\texttt{+},t_{1,b}}^{(0)})$, which gives a new set of completed data that can be used to, again, estimate the model $M_{t_{1,b}}(\boldsymbol{\mu_{t_{1,b}}})$ in Eq. (\ref{eq:LL3_23}). This iterative procedure repeats itself $i$ times until $\hat{n}_{abc,t_{1,b}}^{(i)}$ converges. The resulting set of stabilised completed data are the $\hat{n}_{abc,t_{1,b}}$ in Table \ref{tab:EM}, and they are used to derive maximum likelihood estimates $\hat{m}_{abc,t_{1,b}}$.

The last M-step provides fitted values $\hat{m}_{abc,t}$ for each cell, including the cells with inclusion patterns $001,t_{1}$ and $000,t_{1}$. We refer to these estimates as $\hat{m}_{abc,t}^{\text{NC}}$ and summing up over them for $t=t_{1,b}$ gives a fitted value for $N_{t_{1}}$. We refer to this sum as the nowcast estimate for $N_{t_{1}}$, i.e.
\begin{align} \label{eq:NC}
\hat{N}_{t_{1}}^{\text{NC}} = \sum_{abc \in ABC} \hat{m}_{abc,t_{1,b}}^{\text{NC}}, && 
\end{align}
with $ABC$ the set of all inclusion patterns. In the next section we will use the estimator in Eq. (\ref{eq:NC}) to obtain nowcasts for the number of homeless people in The Netherlands.

A drawback of the EM algorithm, is that is it does not provide the asymptotic variance–covariance matrix \cite[see e.g.][]{Xu2014}, so it seems not possible to derive an analytical expression for the variance of the NC-estimator in Eq.\ \ref{eq:NC}), such as for the regular DSE- and TSE-estimators \cite[see e.g.][Ch. 6, for these expressions]{BishopFienberg1975}. An alternative way to construct confidence intervals for MSE-estimators, proposed by \cite{Buckland1991}, is to construct confidence intervals with the bootstrap approach introduced by \cite{Efron1979}. For the example of homeless people in The Netherlands discussed in the next section, this approach was adopted in \cite{Coumans2017,Zult2023}.

\section{Nowcasting the number of homeless people in The Netherlands} \label{sec:homeless}

In this section we investigate how the MSE nowcasting model performs by using a dataset that is also used to estimate the number of homeless people in The Netherlands, which is discussed in further detail in \cite{Coumans2017}. The estimation procedure is performed annually and is based on three samples that are derived from administrative sources, and we refer to them as sample $A_{y}$, $B_{y}$ and $C_{y}$, where $y$ indicates the year. The resulting TSE estimate for the $1^{st}$ of January of each year is based on a model selection procedure that leads to a TSE model that also includes a set of covariates, namely gender, age, region of stay and region of birth. A more up-to-date time series of estimates of the number of homeless people in The Netherlands, together with their confidence intervals, are presented in \cite{Zult2023}. 

The samples that are used become available during the year, where the first two samples $A_{y}$ and $B_{y}$ are available early during the year, and the third sample $C_{y}$ becomes available somewhere around the third or fourth quarter of the year. These samples are available for each year over the period $2010 - 2023$, except for the year $2019$. Therefore, in $2020$, this sample was replaced with another sample from another administrative source. Because in $2019$ sample $B_{y}$ was missing, for that year the two other samples were also not processed and are therefore also unavailable. The sample size for each sample in each year is presented in Table \ref{tab:sample_size} below.
\begin{table}[h!]
\caption{Sample size for each year} \label{tab:sample_size}
        \begin{tabular}{l|rrr}
                \toprule
                Year & Sample size $A_{y}$ & Sample size $B_{y}$ & Sample size $C_{y}$  \\
                \midrule
                2010 & 2916 & 1746 & 3494 \\
                2011 & 3058 & 1644 & 3812 \\
                2012 & 2594 & 1505 & 3459 \\
                2013 & 2703 & 1491 & 3876 \\
                2014 & 2380 & 1566 & 4267 \\
                2015 & 2232 & 1475 & 4669 \\
                2016 & 2631 & 1130 & 5220 \\
                2017 & 2502 & 1139 & 5611 \\
                2018 & 2456 &  927 & 5824 \\
                2019 & NA & NA & NA \\      
                2020 & 1928 & 2501 & 5808 \\
                2021 & 1992 & 2827 & 6213 \\
                2022 & 2371 & 2263 & 5018 \\
                2023 & 2554 & 3017 & 4315 \\
                \bottomrule
            \end{tabular}
\end{table}

The scheme in which the samples become available implies that at $y = y_{t_{1,b}}$, for the years $2011 - 2018$ and $2021 - 2023$, both a DSE estimate and a NC estimate can be obtained. The fact that a NC estimate, as discussed in Section \ref{sec:combine} and defined in Eq. (\ref{eq:NC}), requires samples from two consecutive years means that it cannot be calculated for the years $2010$, $2019$ and $2020$, because in those years data for the previous or next year are missing.

To simplify the interpretability of the results, both the model selection procedure is simplified by assuming a saturated model, as defined in Eq.\ (\ref{eq:LL3_SAT}), for model $M_{t_{0}}(\boldsymbol{\mu_{t_{1,b}}})$ (see Eq.\ (\ref{eq:LL3_23})). Simply assuming the saturated model in Eq.\ (\ref{eq:LL3_SAT}) for each period, allows for a more fair and straightforward comparison of the resulting estimates between periods. Also, the available covariates are removed from the model by aggregating over the observed frequencies for each covariate. So for example, instead of using the observed frequencies for men and women separately and using gender as an additional explanatory variable in the TSE model equation, the two frequencies are summed-up into one aggregated frequency and gender is left out of the TSE model equation. Aggregating over covariates simplifies the data to the data presented in Table \ref{tab:combined} in Section \ref{sec:combine}. 

To further increase the generality of the analysis, the order in which the samples become available is assumed to be possible in all possible orders. In reality sample $C_{y}$ is available last, but for analytical purposes the order of the samples can be switched without loss of generality. To indicate which samples are assumed to be available first, they are denoted in the subscript. For example, a NC estimate based on sample $A_{y-1}$, $B_{y-1}$, $C_{y-1}$, $A_{y}$ and $B_{y}$, but not $C_{y}$, is denoted as $\hat{N}_{ab,y}^{\text{NC}}$. 

\subsection{Results}
\begin{figure}[h!]
\caption{Estimates of the number of the total number of homeless people in The Netherlands over the periods $2010-2018$ and $2020-2023$.} \label{fig:Homeless_ests}
\vspace{-10pt}
     \centering \includegraphics[scale=0.62]{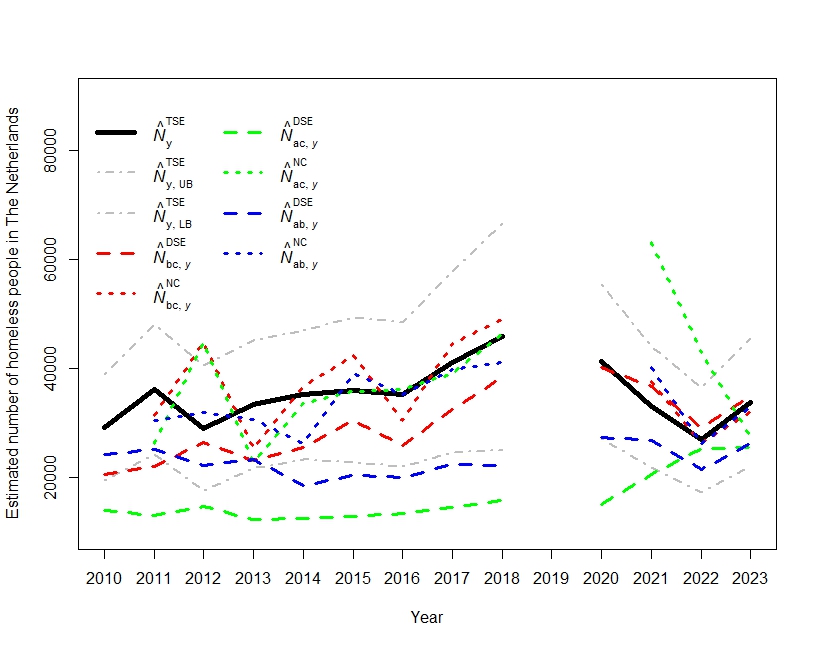}
\end{figure}

Figure \ref{fig:Homeless_ests} presents different time series of population size estimates. It presents the regular TSE estimates $\hat{N}_{y}^{\text{TSE}}$ together with its 95\% confidence interval indicated by $\hat{N}_{y,\text{UB}}^{\text{TSE}}$ and $\hat{N}_{y,\text{LB}}^{\text{TSE}}$. This confidence interval is constructed by using the estimate for the asymptotic variance, for the saturated TSE model, given by \citet[][p. 242]{BishopFienberg1975}. Figure \ref{fig:Homeless_ests} also presents the regular DSE estimates ($\hat{N}_{bc,y}^{\text{DSE}}$, $\hat{N}_{ac,y}^{\text{DSE}}$ and $\hat{N}_{ab,y}^{\text{DSE}}$) and NC estimates ($\hat{N}_{bc,y}^{\text{NC}}$, $\hat{N}_{ac,y}^{\text{NC}}$ and $\hat{N}_{ab,y}^{\text{NC}}$) of the total population size for each year. 
The figure shows that for most years the NC estimates are closer to the regular TSE estimates than to the DSE estimates, which suggest that in this case the nowcasting model assumption of $\lambda_{ab,y}^{AB} = \lambda_{ab,(y-1)}^{AB}$ is more reasonable than the DSE assumption $\lambda_{ab,y}^{AB} = 0$. Furthermore, most NC estimates are within the confidence interval of the regular TSE estimates, while this is not the case for many DSE estimates, especially during the period $2010 - 2018$. In particular the NC estimates $\hat{N}_{ab,y}^{\text{NC}}$ represented by the blue (short-dotted) line, which represent the real situation, seems to perform reasonably well in most years. The red and green (short-dotted) line generally perform a bit less well, in particular the estimates $\hat{N}_{ac,y}^{\text{NC}}$ for the years $2021$ and $2022$ are quite bad. Interestingly, although not shown here, these two outliers disappear when for $M_{t_{0}}(\boldsymbol{\mu_{t_{1,b}}})$, for the years $2021$ and $2022$ instead of the saturated model in Eq.\ (\ref{eq:LL3_SAT}), the two-pair dependence model in Eq.\ (\ref{eq:LL3_2PD_I}) is chosen. This suggests that by applying a well-designed model selection procedure in the nowcasting approach as in \cite{Coumans2017}, which is not possible in DSE, the accuracy of the nowcasting estimates in Figure \ref{fig:Homeless_ests} can be further improved.

Figure \ref{fig:Homeless_ests} also shows that the lagged TSE estimates, i.e. the TSE estimates of the previous year, are also within the TSE confidence intervals. This raises the question whether in this case it makes any sense to apply the nowcasting model. To look deeper into this issue, Table \ref{tab:compare_TSE} presents the annual differences between the TSE estimates and the lagged TSE estimates (presented in the column $\hat{N}_{(y-1)}^{\text{TSE}}-\hat{N}_{y}^{\text{TSE}}$), together with the annual differences between the TSE estimates and the three different NC estimates.
\begin{table}[h!]
\caption{Difference per year ($\times 1000$) between the TSE estimate and different estimates for each year} \label{tab:compare_TSE}
        \begin{tabular}{l|r@{}l|r@{}lr@{}lr@{}l}
                \toprule
                Year & \multicolumn{2}{l|}{$\hat{N}_{(y-1)}^{\text{TSE}}-\hat{N}_{y}^{\text{TSE}}$} & \multicolumn{2}{l}{$\hat{N}_{bc,y}^{\text{NC}}-\hat{N}_{y}^{\text{TSE}}$} & \multicolumn{2}{l}{$\hat{N}_{ac,y}^{\text{NC}}-\hat{N}_{y}^{\text{TSE}}$} & \multicolumn{2}{l}{$\hat{N}_{ab,y}^{\text{NC}}-\hat{N}_{y}^{\text{TSE}}$}  \\
                \midrule
                2011 & -6&.9 & -4&.8 &  -9&.9 & -5&.7 \\
                2012 & -7&.0 & 15&.7 &  15&.6 &  2&.8 \\
                2013 &  4&.3 & -8&.1 & -10&.7 & -2&.8 \\
                2014 &  1&.8 &  1&.6 &  -1&.6 & -8&.9 \\
                2015 &  0&.9 &  6&.3 &  -0&.2 &  3&.1 \\
                2016 & -0&.8 & -4&.7 &   0&.9 &  0&.0 \\
                2017 &  6&.0 &  3&.5 &  -2&.2 & -1&.4 \\
                2018 &  4&.6 &  3&.2 &   0&.6 & -4&.7 \\
                2021 & -8&.3 &  4&.6 &  30&.2 &  7&.2 \\
                2022 & -6&.0 & -0&.7 &  16&.1 & -0&.6 \\
                2023 &  6&.9 & -1&.8 &  -6&.5 & -0&.9 \\
                \midrule
                MAE &  4&.5 &  4&.7 &   8&.1 &  3&.3 \\
                \bottomrule
            \end{tabular}
\end{table}
Table \ref{tab:compare_TSE} shows that the differences between NC estimates and TSE estimates clearly differ for each sample order. The best results are in the last column $\hat{N}_{ab,y}^{\text{NC}}-\hat{N}_{y}^{\text{TSE}}$, which has the lowest mean absolute error (MAE, $3.3$) and implies that in the case of the homeless data the nowcasting model with sample $C_{y}$ being later (which corresponds to the real situation), gives the best nowcasting results that are also slightly more accurate than the lagged TSE estimates. To further compare the predictive accuracy of the lagged TSE estimate and the three NC estimates, we also apply the Diebold-Mariano test \cite[][]{Diebold1995}, which can be used to tests for statistical differences in predictive accuracy of different series. For the 95\% confidence level, this test does not reject statistically significant differences in predictive accuracy between the four different series, which may also be due to the limited number of estimates over time.

\begin{figure}[h!]
\caption{Observations and nowcasts of the number of homeless people that are uniquely observed in the last available sample over the periods 2010-2018 and 2020-2023.} \label{fig:Homeless_n001}
\vspace{-10pt}
     \centering \includegraphics[scale=0.62]{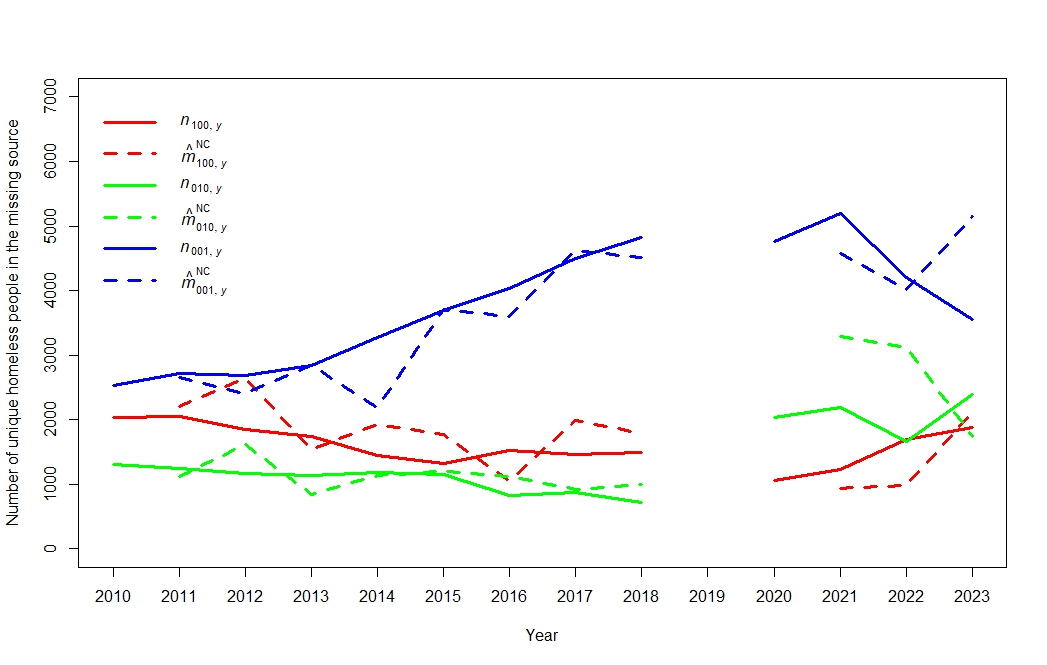}
\end{figure}

An advantage that is common to nowcasting models that can also be utilised in the study presented in this section, is that NC estimates can often be compared with true values that become available later. In most MSE applications such an evaluation is not possible, simply because the true population size usually remains unknown. However, in this study the nowcasting model, additional to the NC estimate for the total population, it also provides a NC estimate $\hat{m}_{001,y}^{\text{NC}}$, for which an observation $n_{001,y}$ becomes available when the last sample becomes available. This comparison is presented in Figure \ref{fig:Homeless_n001}.

Figure \ref{fig:Homeless_n001} shows, as solid lines, the observed counts $n_{100,y}$, $n_{010,y}$ and $n_{001,y}$, and as dotted lines their corresponding NC estimates $\hat{m}_{100,y}^{\text{NC}},$ $\hat{m}_{010,y}^{\text{NC}}$ and $\hat{m}_{001,y}^{\text{NC}}$. In each NC estimate, it was assumed that the sample indicated by a \q{$1$} in the inclusion pattern in the subscript, was not yet available.  For example, $\hat{m}_{001,y}^{\text{NC}}$ is a nowcast that is based on the samples $A_{y-1}$, $B_{y-1}$, $C_{y-1}$, $A_{y}$ and $B_{y}$, and not $C_{y}$. Figure \ref{fig:Homeless_n001} shows that irrespective of which sample is missing, over the period $2010 - 2018$, the nowcasting model estimates $\hat{m}_{100,y}^{\text{NC}}$, $\hat{m}_{010,y}^{\text{NC}}$ and $\hat{m}_{001,y}^{\text{NC}}$ follow a similar trend as the (later) observed counts $n_{100}$, $n_{010}$ and $n_{001}$. In particular for the red and green lines this trend becomes less clear after $2020$. This can be explained by the fact that after $2019$, sample $B_{y}$ was a sample from a different administrative source than before. Before $2019$ sample $B_{y}$ was a sample of homeless people who suffered from drug addiction problems and after $2019$ sample $B_{y}$ was a sample of homeless people who were ex-prisoners who received reintegration support.

Just like in Figure \ref{fig:Homeless_ests}, for the estimates presented in Figure \ref{fig:Homeless_n001}, one may also suspect that it might be easier and more accurate to simply use the lagged observation instead of the NC estimate (e.g. use $n_{001,y-1}$ instead of $\hat{m}_{001,y}^{\text{NC}}$ as an estimate for $m_{001,y}$). When the MAEs (not provided here) of these lagged observations are compared with the MAEs of the NC estimates in Figure \ref{fig:Homeless_n001}, this indeed shows to be the case for the most stable series $n_{100,y}$ and $n_{010,y}$. However, we see that for the least stable series $n_{001,y}$, the difference in MAE of the lagged observations and NC estimates is negligible. This underlines the notion that the MSE nowcasting model becomes more valuable in the case of less stable population size estimates.

Both Figure \ref{fig:Homeless_ests} and \ref{fig:Homeless_n001} show that for some years and for some orders of sample availability the quality of the nowcast is better than for other years. Here it is interesting to note that both Figures indicate that the NC estimates are most accurate when sample $C_{y}$ is missing. This is somewhat surprising, because as Table \ref{tab:sample_size} shows, sample $C_{y}$ is also the largest sample in each year. An alternative reason for differences between NC and TSE estimates can be found in the stability of the pairwise-dependence parameter $\lambda_{ab,y}^{AB}$, as introduced in Eq.\ (\ref{eq:LL2}), between two consecutive years.
Note that this parameter can be estimated when the third sample for that year is available. Figure \ref{fig:lambda} below shows these estimates for $\lambda_{ab,y}^{AB}$, $\lambda_{ac,y}^{AC}$ and $\lambda_{bc,y}^{BC}$, over time. 
\begin{figure}[ht!]
\caption{Coefficient estimates of $\mu_{ab,y}^{AB}$, $\mu_{ac,y}^{AC}$ and $\mu_{bc,y}^{BC}$ over the periods $2011-2018$ and $2021-2023$.} \label{fig:lambda}
\vspace{-10pt}
     \centering \includegraphics[scale=0.62]{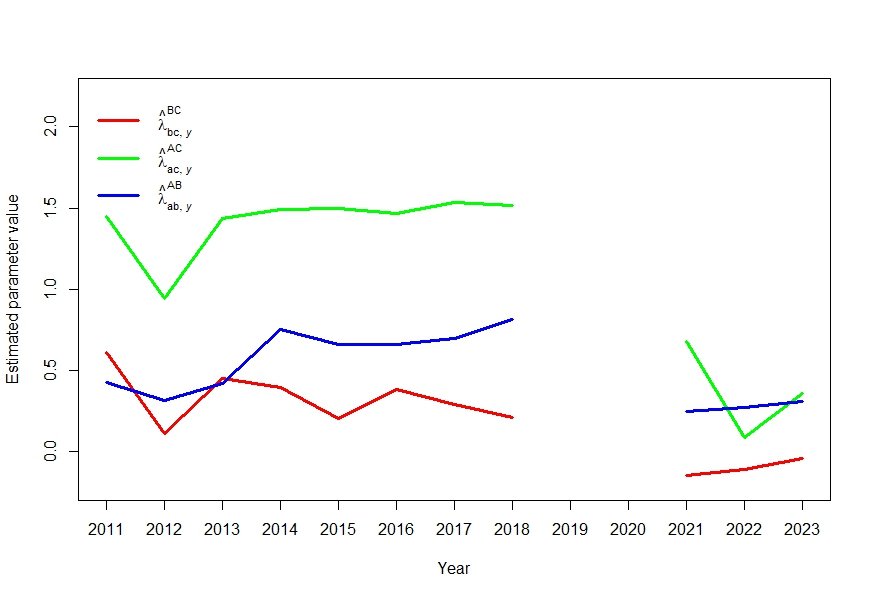}
\end{figure}

Figure \ref{fig:lambda} clearly shows three separate time series, which indicates that there is at least some stability in $\lambda_{ab,y}^{AB}$, $\lambda_{ac,y}^{AC}$ and $\lambda_{bc,y}^{BC}$ over time. Figure \ref{fig:lambda} also shows that the estimated pairwise-dependencies became closer to zero when sample $B_{y}$ was replaced after $2019$. This proximity to zero explains the larger accuracy of most of the DSE estimates in the years $2020-2023$, as was shown in Figure \ref{fig:Homeless_ests}. Ideally, each time series of estimates presented in Figure \ref{fig:lambda} are completely stable between periods. Long-term stability can in theory be tested with, for example, the well-known Dickey-Fuller unit root test \citep{Dickey1979}. However, this type of test requires the availability of a (much) longer time series to be reliable. A simple look at Figure \ref{fig:lambda} shows that stability between two consecutive years is not always the case, for example for $\hat{\lambda}_{bc,y}^{BC}$ in $2012$ and $2013$, and for $\hat{\lambda}_{ac,y}^{AC}$ in $2012$, $2013$, $2022$ and $2023$. We do not have an explanation for these shifts, but it is interesting to compare them to the time series in Figure \ref{fig:Homeless_ests} and the annual errors in Table \ref{tab:compare_TSE}. This comparisons shows that these larger shifts in $\hat{\lambda}_{bc,y}^{BC}$ and $\hat{\lambda}_{ac,y}^{AC}$ correspond to the years with larger nowcasting errors. This says that the stability of the pairwise-dependence parameter can be more important than the size of the missing sample.

\section{Discussion}
In this paper we propose to combine dual- and triple system estimation over two periods by means of the expectation-maximisation algorithm to obtain a preliminary estimate, that we have coined an MSE nowcast estimate. The advantage of this approach is that, in the presence of older samples, it allows estimation with two (early available) samples, like in DSE. The difference with regular DSE is that the independence assumption in DSE can be replaced by a more relaxed assumption, i.e., the pairwise-dependence of the two samples equals the corresponding pairwise-dependence in the previous period. Whether the NC estimates are accurate depends on the validity of this assumption. 

To test for the stability of pairwise-dependencies over time, a formal statistical test such as a Dickey-Fuller unit root test \citep{Dickey1979}, usually requires longer time series than that are available in most TSE applications, including the example on the number of homeless people in The Netherlands that was presented in Section \ref{sec:homeless}. Therefore, in this example, the accuracy of the NC estimates was investigated in four other ways. First, the NC estimates are compared to the regular DSE estimates, which indicated that in our example, for most years, the nowcast assumption of stable pairwise-dependencies is more realistic than the DSE assumption of pairwise-independence. Second, we checked whether the NC estimates were within the confidence intervals of the TSE estimates, which was mostly the case. Third, the NC estimates of $m_{001,y}$ were compared to the corresponding observations $n_{001,y}$. Such a comparison is common in time series analysis, but rare in TSE, where population sizes usually remain unknown. This comparison showed that the estimates for $m_{001,y}$ and $n_{001,y}$ follow similar trends and are reasonable alike for most years, although for some years there are substantial differences. Finally, we estimated the pairwise-dependence parameters for the first two samples over time, which for each years becomes possible when the third sample is available. This analysis showed that for most years, the pairwise-dependence parameter estimates are reasonably stable, but not for each year. These unstable years directly correspond to larger differences between NC estimates and TSE estimates.

We conclude that our method to obtain NC estimates works and can be useful under two conditions. First, there needs to be some volatility in the population, because otherwise it may be easier and more accurate to simply use the population size estimate of the previous period. Second, the pairwise-dependence parameters of the first two samples needs to be reasonably stable. If this is not the case, the NC estimates can become seriously inaccurate, although they may still be more accurate than regular DSE estimates.

\bibliography{references.bib}

\begin{thebibliography}{}

\bibitem [\protect \citeauthoryear {%
Baffour%
, Brown%
\BCBL {}\ \BBA {} Smith%
}{%
Baffour%
\ \protect \BOthers {.}}{%
{\protect \APACyear {2013}}%
}]{%
Baffour2013}
\APACinsertmetastar {%
Baffour2013}%
\begin{APACrefauthors}%
Baffour, B.%
, Brown, J\BPBI J.%
\BCBL {}\ \BBA {} Smith, P\BPBI W.%
\end{APACrefauthors}%
\unskip\
\newblock
\APACrefYearMonthDay{2013}{}{}.
\newblock
{\BBOQ}\APACrefatitle {An investigation of triple system estimators in censuses} {An investigation of triple system estimators in censuses}.{\BBCQ}
\newblock
\APACjournalVolNumPages{Statistical Journal of the IAOS}{29}{}{53--68}.
\newblock
\begin{APACrefURL} \url{https://doi.org/10.3233/SJI-130760} \end{APACrefURL}
\newblock
\begin{APACrefDOI} \doi{10.3233/SJI-130760} \end{APACrefDOI}
\PrintBackRefs{\CurrentBib}

\bibitem [\protect \citeauthoryear {%
Baffour%
, Brown%
\BCBL {}\ \BBA {} Smith%
}{%
Baffour%
\ \protect \BOthers {.}}{%
{\protect \APACyear {2021}}%
}]{%
Baffour2021}
\APACinsertmetastar {%
Baffour2021}%
\begin{APACrefauthors}%
Baffour, B.%
, Brown, J\BPBI J.%
\BCBL {}\ \BBA {} Smith, P\BPBI W.%
\end{APACrefauthors}%
\unskip\
\newblock
\APACrefYearMonthDay{2021}{}{}.
\newblock
{\BBOQ}\APACrefatitle {Latent Class Analysis for Estimating an Unknown Population Size – with Application to Censuses} {Latent class analysis for estimating an unknown population size – with application to censuses}.{\BBCQ}
\newblock
\APACjournalVolNumPages{Journal of Official Statistics}{37}{3}{673-697}.
\newblock
\begin{APACrefDOI} \doi{10.2478/jos-2021-0030} \end{APACrefDOI}
\PrintBackRefs{\CurrentBib}

\bibitem [\protect \citeauthoryear {%
Bird%
\ \BBA {} King%
}{%
Bird%
\ \BBA {} King%
}{%
{\protect \APACyear {2018}}%
}]{%
Bird2018}
\APACinsertmetastar {%
Bird2018}%
\begin{APACrefauthors}%
Bird, S\BPBI M.%
\BCBT {}\ \BBA {} King, R.%
\end{APACrefauthors}%
\unskip\
\newblock
\APACrefYearMonthDay{2018}{}{}.
\newblock
{\BBOQ}\APACrefatitle {Multiple Systems Estimation (or Capture-Recapture Estimation) to Inform Public Policy.} {Multiple systems estimation (or capture-recapture estimation) to inform public policy.}{\BBCQ}
\newblock
\APACjournalVolNumPages{Annual review of statistics and its application}{5}{}{95–-118}.
\newblock
\begin{APACrefDOI} \doi{10.1146/annurev-statistics-031017-100641} \end{APACrefDOI}
\PrintBackRefs{\CurrentBib}

\bibitem [\protect \citeauthoryear {%
Bishop%
, Fienberg%
\BCBL {}\ \BBA {} Holland%
}{%
Bishop%
\ \protect \BOthers {.}}{%
{\protect \APACyear {1975}}%
}]{%
BishopFienberg1975}
\APACinsertmetastar {%
BishopFienberg1975}%
\begin{APACrefauthors}%
Bishop, Y\BPBI M\BPBI M.%
, Fienberg, S\BPBI E.%
\BCBL {}\ \BBA {} Holland, P\BPBI W.%
\end{APACrefauthors}%
\unskip\
\newblock
\APACrefYear{1975}.
\newblock
\APACrefbtitle {Discrete Multivariate Analysis} {Discrete multivariate analysis}.
\newblock
\APACaddressPublisher{}{Springer New York, NY}.
\newblock
\begin{APACrefURL} \url{https://link.springer.com/book/10.1007/978-0-387-72806-3} \end{APACrefURL}
\newblock
\APACrefnote{\href{https://esl.hohoweiya.xyz/references/Discrete-Multivariate-Analysis.pdf}{link 10.1007/978-0-387-72806-3}}
\newblock
\begin{APACrefDOI} \doi{10.1007/978-0-387-72806-3} \end{APACrefDOI}
\PrintBackRefs{\CurrentBib}

\bibitem [\protect \citeauthoryear {%
Buckland%
\ \BBA {} Garthwaite%
}{%
Buckland%
\ \BBA {} Garthwaite%
}{%
{\protect \APACyear {1991}}%
}]{%
Buckland1991}
\APACinsertmetastar {%
Buckland1991}%
\begin{APACrefauthors}%
Buckland, S\BPBI T.%
\BCBT {}\ \BBA {} Garthwaite, P\BPBI H.%
\end{APACrefauthors}%
\unskip\
\newblock
\APACrefYearMonthDay{1991}{}{}.
\newblock
{\BBOQ}\APACrefatitle {Quantifying Precision of Mark-Recapture Estimates Using the Bootstrap and Related Methods} {Quantifying precision of mark-recapture estimates using the bootstrap and related methods}.{\BBCQ}
\newblock
\APACjournalVolNumPages{Biometrics}{47}{1}{255--268}.
\newblock
\begin{APACrefURL} [{2024-10-15}]\url{http://www.jstor.org/stable/2532510} \end{APACrefURL}
\PrintBackRefs{\CurrentBib}

\bibitem [\protect \citeauthoryear {%
Coumans%
, Cruyff%
, van~der Heijden%
, Wolf%
\BCBL {}\ \BBA {} Schmeets%
}{%
Coumans%
\ \protect \BOthers {.}}{%
{\protect \APACyear {2017}}%
}]{%
Coumans2017}
\APACinsertmetastar {%
Coumans2017}%
\begin{APACrefauthors}%
Coumans, M\BPBI A.%
, Cruyff, M.%
, van~der Heijden, P\BPBI G\BPBI M.%
, Wolf, J.%
\BCBL {}\ \BBA {} Schmeets, H.%
\end{APACrefauthors}%
\unskip\
\newblock
\APACrefYearMonthDay{2017}{}{}.
\newblock
{\BBOQ}\APACrefatitle {Estimating Homelessness in {The Netherlands} Using a Capture-Recapture Approach} {Estimating homelessness in {The Netherlands} using a capture-recapture approach}.{\BBCQ}
\newblock
\APACjournalVolNumPages{Social Indicators Research}{130}{1}{89--212}.
\newblock
\begin{APACrefURL} \url{https://doi.org/10.1007/s11205-015-1171-7} \end{APACrefURL}
\newblock
\begin{APACrefDOI} \doi{10.1007/s11205-015-1171-7} \end{APACrefDOI}
\PrintBackRefs{\CurrentBib}

\bibitem [\protect \citeauthoryear {%
Dempster%
, Laird%
\BCBL {}\ \BBA {} Rubin%
}{%
Dempster%
\ \protect \BOthers {.}}{%
{\protect \APACyear {1977}}%
}]{%
Dempster1977}
\APACinsertmetastar {%
Dempster1977}%
\begin{APACrefauthors}%
Dempster, A\BPBI P.%
, Laird, N\BPBI M.%
\BCBL {}\ \BBA {} Rubin, D\BPBI B.%
\end{APACrefauthors}%
\unskip\
\newblock
\APACrefYearMonthDay{1977}{}{}.
\newblock
{\BBOQ}\APACrefatitle {Maximum Likelihood from Incomplete Data via the {EM} Algorithm} {Maximum likelihood from incomplete data via the {EM} algorithm}.{\BBCQ}
\newblock
\APACjournalVolNumPages{Journal of the Royal Statistical Society. Series B (Methodological)}{39}{1}{1--38}.
\newblock
\begin{APACrefURL} [{2024-03-21}]\url{http://www.jstor.org/stable/2984875} \end{APACrefURL}
\PrintBackRefs{\CurrentBib}

\bibitem [\protect \citeauthoryear {%
Dickey%
\ \BBA {} Fuller%
}{%
Dickey%
\ \BBA {} Fuller%
}{%
{\protect \APACyear {1979}}%
}]{%
Dickey1979}
\APACinsertmetastar {%
Dickey1979}%
\begin{APACrefauthors}%
Dickey, D\BPBI A.%
\BCBT {}\ \BBA {} Fuller, W\BPBI A.%
\end{APACrefauthors}%
\unskip\
\newblock
\APACrefYearMonthDay{1979}{}{}.
\newblock
{\BBOQ}\APACrefatitle {Distribution of the Estimators for Autoregressive Time Series with a Unit Root} {Distribution of the estimators for autoregressive time series with a unit root}.{\BBCQ}
\newblock
\APACjournalVolNumPages{Journal of the American Statistical Association}{74}{366a}{427--431}.
\newblock
\begin{APACrefDOI} \doi{10.1080/01621459.1979.10482531} \end{APACrefDOI}
\PrintBackRefs{\CurrentBib}

\bibitem [\protect \citeauthoryear {%
Diebold%
\ \BBA {} Mariano%
}{%
Diebold%
\ \BBA {} Mariano%
}{%
{\protect \APACyear {1995}}%
}]{%
Diebold1995}
\APACinsertmetastar {%
Diebold1995}%
\begin{APACrefauthors}%
Diebold, F\BPBI X.%
\BCBT {}\ \BBA {} Mariano, R\BPBI S.%
\end{APACrefauthors}%
\unskip\
\newblock
\APACrefYearMonthDay{1995}{}{}.
\newblock
{\BBOQ}\APACrefatitle {Comparing Predictive Accuracy} {Comparing predictive accuracy}.{\BBCQ}
\newblock
\APACjournalVolNumPages{Journal of Business \& Economic Statistics}{13}{3}{253--263}.
\newblock
\begin{APACrefURL} \url{https://www.tandfonline.com/doi/abs/10.1080/07350015.1995.10524599} \end{APACrefURL}
\newblock
\begin{APACrefDOI} \doi{10.1080/07350015.1995.10524599} \end{APACrefDOI}
\PrintBackRefs{\CurrentBib}

\bibitem [\protect \citeauthoryear {%
Efron%
}{%
Efron%
}{%
{\protect \APACyear {1979}}%
}]{%
Efron1979}
\APACinsertmetastar {%
Efron1979}%
\begin{APACrefauthors}%
Efron, B.%
\end{APACrefauthors}%
\unskip\
\newblock
\APACrefYearMonthDay{1979}{}{}.
\newblock
{\BBOQ}\APACrefatitle {Bootstrap Methods: Another Look at the Jackknife} {Bootstrap methods: Another look at the jackknife}.{\BBCQ}
\newblock
\APACjournalVolNumPages{The Annals of Statistics}{7}{1}{1--26}.
\newblock
\begin{APACrefURL} [{2024-10-15}]\url{http://www.jstor.org/stable/2958830} \end{APACrefURL}
\PrintBackRefs{\CurrentBib}

\bibitem [\protect \citeauthoryear {%
Fienberg%
}{%
Fienberg%
}{%
{\protect \APACyear {1972}}%
}]{%
Fienberg1972}
\APACinsertmetastar {%
Fienberg1972}%
\begin{APACrefauthors}%
Fienberg, S\BPBI E.%
\end{APACrefauthors}%
\unskip\
\newblock
\APACrefYearMonthDay{1972}{}{}.
\newblock
{\BBOQ}\APACrefatitle {The Multiple Recapture Census for Closed Populations and Incomplete $2^k$ Contingency Tables} {The multiple recapture census for closed populations and incomplete $2^k$ contingency tables}.{\BBCQ}
\newblock
\APACjournalVolNumPages{Biometrika}{59}{3}{591--603}.
\newblock
\begin{APACrefURL} \url{https://doi.org/10.2307/2334810} \end{APACrefURL}
\newblock
\begin{APACrefDOI} \doi{10.2307/2334810} \end{APACrefDOI}
\PrintBackRefs{\CurrentBib}

\bibitem [\protect \citeauthoryear {%
Gerritse%
, Bakker%
, de Wolf%
\BCBL {}\ \BBA {} van~der Heijden%
}{%
Gerritse%
\ \protect \BOthers {.}}{%
{\protect \APACyear {2016}}%
}]{%
Gerritse2016}
\APACinsertmetastar {%
Gerritse2016}%
\begin{APACrefauthors}%
Gerritse, S\BPBI C.%
, Bakker, B\BPBI F\BPBI M.%
, de Wolf, P.%
\BCBL {}\ \BBA {} van~der Heijden, P\BPBI G\BPBI M.%
\end{APACrefauthors}%
\unskip\
\newblock
\APACrefYearMonthDay{2016}{}{}.
\newblock
{\BBOQ}\APACrefatitle {Under coverage of the population register in {The Netherlands}} {Under coverage of the population register in {The Netherlands}}.{\BBCQ}
\newblock
\begin{APACrefURL} \url{https://dspace.library.uu.nl/bitstream/handle/1874/356071/register.pdf?sequence=1} \end{APACrefURL}
\PrintBackRefs{\CurrentBib}

\bibitem [\protect \citeauthoryear {%
Lincoln%
}{%
Lincoln%
}{%
{\protect \APACyear {1930}}%
}]{%
Lincoln1930}
\APACinsertmetastar {%
Lincoln1930}%
\begin{APACrefauthors}%
Lincoln, F\BPBI C.%
\end{APACrefauthors}%
\unskip\
\newblock
\APACrefYear{1930}.
\newblock
\APACrefbtitle {Calculating Waterfowl Abundance on the Basis of Banding Returns} {Calculating waterfowl abundance on the basis of banding returns}\ (\BVOL~118).
\newblock
\APACaddressPublisher{}{United States Department of Agriculture}.
\newblock
\begin{APACrefURL} \url{https://doi.org/10.5962/bhl.title.64010} \end{APACrefURL}
\newblock
\begin{APACrefDOI} \doi{10.5962/bhl.title.64010} \end{APACrefDOI}
\PrintBackRefs{\CurrentBib}

\bibitem [\protect \citeauthoryear {%
Lindsay%
}{%
Lindsay%
}{%
{\protect \APACyear {1995}}%
}]{%
Lindsay1995}
\APACinsertmetastar {%
Lindsay1995}%
\begin{APACrefauthors}%
Lindsay, B\BPBI G.%
\end{APACrefauthors}%
\unskip\
\newblock
\APACrefYearMonthDay{1995}{}{}.
\newblock
{\BBOQ}\APACrefatitle {Mixture Models: Theory, Geometry and Applications} {Mixture models: Theory, geometry and applications}.{\BBCQ}
\newblock
\APACjournalVolNumPages{NSF-CBMS Regional Conference Series in Probability and Statistics}{5}{}{i--163}.
\newblock
\begin{APACrefURL} [{2024-03-22}]\url{http://www.jstor.org/stable/4153184} \end{APACrefURL}
\PrintBackRefs{\CurrentBib}

\bibitem [\protect \citeauthoryear {%
Petersen%
}{%
Petersen%
}{%
{\protect \APACyear {1896}}%
}]{%
Petersen1896}
\APACinsertmetastar {%
Petersen1896}%
\begin{APACrefauthors}%
Petersen, C\BPBI G\BPBI J.%
\end{APACrefauthors}%
\unskip\
\newblock
\APACrefYearMonthDay{1896}{}{}.
\newblock
{\BBOQ}\APACrefatitle {The Yearly Immigration of Young Plaice Into the {Limfjord} From the {German Sea}} {The yearly immigration of young plaice into the {Limfjord} from the {German Sea}}.{\BBCQ}
\newblock
\APACjournalVolNumPages{Report of the Danish Biological Station}{6}{}{5--84}.
\newblock
\begin{APACrefURL} \url{https://archive.org/details/reportofdanishbi06dans/page/n1/mode/2up} \end{APACrefURL}
\PrintBackRefs{\CurrentBib}

\bibitem [\protect \citeauthoryear {%
Seber%
}{%
Seber%
}{%
{\protect \APACyear {1982}}%
}]{%
Seber1982}
\APACinsertmetastar {%
Seber1982}%
\begin{APACrefauthors}%
Seber, G\BPBI A\BPBI F.%
\end{APACrefauthors}%
\unskip\
\newblock
\APACrefYear{1982}.
\newblock
\APACrefbtitle {The Estimation of Animal Abundance and Related Parameters} {The estimation of animal abundance and related parameters}\ (\PrintOrdinal{Second}\ \BEd).
\newblock
\APACaddressPublisher{}{London: Griffin}.
\newblock
\begin{APACrefURL} \url{https://archive.org/details/estimationofanim0000sebe/page/n5/mode/2up} \end{APACrefURL}
\PrintBackRefs{\CurrentBib}

\bibitem [\protect \citeauthoryear {%
van~der Heijden%
\ \protect \BOthers {.}}{%
van~der Heijden%
\ \protect \BOthers {.}}{%
{\protect \APACyear {2021}}%
}]{%
Heijden2020}
\APACinsertmetastar {%
Heijden2020}%
\begin{APACrefauthors}%
van~der Heijden, P\BPBI G\BPBI M.%
, Cruyff, M.%
, Smith, P\BPBI A.%
, Bycroft, C.%
, Graham, P.%
\BCBL {}\ \BBA {} {Matheson-Dunning}, N.%
\end{APACrefauthors}%
\unskip\
\newblock
\APACrefYearMonthDay{2021}{}{}.
\newblock
{\BBOQ}\APACrefatitle {Multiple System Estimation using Covariates having Missing Values and Measurement Error: Estimating the Size of the {Māori} Population in {New Zealand}} {Multiple system estimation using covariates having missing values and measurement error: Estimating the size of the {Māori} population in {New Zealand}}.{\BBCQ}
\newblock
\APACjournalVolNumPages{Journal of the Royal Statistical Society Series A: Statistics in Society}{185}{1}{156--177}.
\newblock
\begin{APACrefURL} \url{https://doi.org/10.1111/rssa.12731} \end{APACrefURL}
\newblock
\begin{APACrefDOI} \doi{10.1111/rssa.12731} \end{APACrefDOI}
\PrintBackRefs{\CurrentBib}

\bibitem [\protect \citeauthoryear {%
van~der Heijden%
, Whittaker%
, Cruyff%
, Bakker%
\BCBL {}\ \BBA {} van~der Vliet%
}{%
van~der Heijden%
\ \protect \BOthers {.}}{%
{\protect \APACyear {2012}}%
}]{%
vdHeijden2012}
\APACinsertmetastar {%
vdHeijden2012}%
\begin{APACrefauthors}%
van~der Heijden, P\BPBI G\BPBI M.%
, Whittaker, J.%
, Cruyff, M.%
, Bakker, B\BPBI F\BPBI M.%
\BCBL {}\ \BBA {} van~der Vliet, R.%
\end{APACrefauthors}%
\unskip\
\newblock
\APACrefYearMonthDay{2012}{}{}.
\newblock
{\BBOQ}\APACrefatitle {People born in the Middle East but residing in the Netherlands: Invariant population size estimates and the role of active and passive covariates} {People born in the middle east but residing in the netherlands: Invariant population size estimates and the role of active and passive covariates}.{\BBCQ}
\newblock
\APACjournalVolNumPages{The Annals of Applied Statistics}{6}{3}{831--852}.
\newblock
\APACrefnote{\href{https://doi.org/10.1214/12-AOAS536}{link}}
\newblock
\begin{APACrefDOI} \doi{10.1214/12-AOAS536} \end{APACrefDOI}
\PrintBackRefs{\CurrentBib}

\bibitem [\protect \citeauthoryear {%
Wolter%
}{%
Wolter%
}{%
{\protect \APACyear {1986}}%
}]{%
Wolter1986}
\APACinsertmetastar {%
Wolter1986}%
\begin{APACrefauthors}%
Wolter, K\BPBI M.%
\end{APACrefauthors}%
\unskip\
\newblock
\APACrefYearMonthDay{1986}{}{}.
\newblock
{\BBOQ}\APACrefatitle {Some Coverage Error Models for Census Data} {Some coverage error models for census data}.{\BBCQ}
\newblock
\APACjournalVolNumPages{Journal of the American Statistical Association}{81}{}{338--346}.
\newblock
\begin{APACrefURL} \url{https://doi.org/10.2307/2289222} \end{APACrefURL}
\newblock
\begin{APACrefDOI} \doi{10.2307/2289222} \end{APACrefDOI}
\PrintBackRefs{\CurrentBib}

\bibitem [\protect \citeauthoryear {%
Xu%
, Baines%
\BCBL {}\ \BBA {} Wang%
}{%
Xu%
\ \protect \BOthers {.}}{%
{\protect \APACyear {2014}}%
}]{%
Xu2014}
\APACinsertmetastar {%
Xu2014}%
\begin{APACrefauthors}%
Xu, C.%
, Baines, P\BPBI D.%
\BCBL {}\ \BBA {} Wang, J\BHBI L.%
\end{APACrefauthors}%
\unskip\
\newblock
\APACrefYearMonthDay{2014}{}{}.
\newblock
{\BBOQ}\APACrefatitle {Standard error estimation using the {EM} algorithm for the joint modeling of survival and longitudinal data.} {Standard error estimation using the {EM} algorithm for the joint modeling of survival and longitudinal data.}{\BBCQ}
\newblock
\APACjournalVolNumPages{Biostatistics}{15}{4}{731--44}.
\newblock
\begin{APACrefURL} \url{https://doi.org/10.1093/biostatistics/kxu015} \end{APACrefURL}
\newblock
\begin{APACrefDOI} \doi{10.1093/biostatistics/kxu015} \end{APACrefDOI}
\PrintBackRefs{\CurrentBib}

\bibitem [\protect \citeauthoryear {%
Zult%
, van~der Heijden%
\BCBL {}\ \BBA {} Bakker%
}{%
Zult%
\ \protect \BOthers {.}}{%
{\protect \APACyear {2023}}%
}]{%
Zult2023}
\APACinsertmetastar {%
Zult2023}%
\begin{APACrefauthors}%
Zult, D\BPBI B.%
, van~der Heijden, P\BPBI G\BPBI M.%
\BCBL {}\ \BBA {} Bakker, B\BPBI F\BPBI M.%
\end{APACrefauthors}%
\unskip\
\newblock
\APACrefYearMonthDay{2023}{}{}.
\newblock
{\BBOQ}\APACrefatitle {Bias correction in multiple-systems estimation} {Bias correction in multiple-systems estimation}.{\BBCQ}
\newblock
\APACjournalVolNumPages{arXiv}{}{}{}.
\newblock
\begin{APACrefURL} \url{https://doi.org/10.48550/arXiv.2311.01297} \end{APACrefURL}
\PrintBackRefs{\CurrentBib}

\bibitem [\protect \citeauthoryear {%
Zwane%
\ \BBA {} van~der Heijden%
}{%
Zwane%
\ \BBA {} van~der Heijden%
}{%
{\protect \APACyear {2007}}%
}]{%
Zwane2007}
\APACinsertmetastar {%
Zwane2007}%
\begin{APACrefauthors}%
Zwane, E\BPBI N.%
\BCBT {}\ \BBA {} van~der Heijden, P\BPBI G\BPBI M.%
\end{APACrefauthors}%
\unskip\
\newblock
\APACrefYearMonthDay{2007}{}{}.
\newblock
{\BBOQ}\APACrefatitle {Analysing capture–recapture data when some variables of heterogeneous catchability are not collected or asked in all registrations} {Analysing capture–recapture data when some variables of heterogeneous catchability are not collected or asked in all registrations}.{\BBCQ}
\newblock
\APACjournalVolNumPages{Statistics in Medicine}{26}{}{1069--1089}.
\newblock
\begin{APACrefURL} \url{https://doi.org/10.1002/sim.2577} \end{APACrefURL}
\newblock
\begin{APACrefDOI} \doi{10.1002/sim.2577} \end{APACrefDOI}
\PrintBackRefs{\CurrentBib}

\bibitem [\protect \citeauthoryear {%
Zwane%
, van der Pal-de Bruin%
\BCBL {}\ \BBA {} van~der Heijden%
}{%
Zwane%
\ \protect \BOthers {.}}{%
{\protect \APACyear {2004}}%
}]{%
Zwane2004}
\APACinsertmetastar {%
Zwane2004}%
\begin{APACrefauthors}%
Zwane, E\BPBI N.%
, van der Pal-de Bruin, K.%
\BCBL {}\ \BBA {} van~der Heijden, P\BPBI G\BPBI M.%
\end{APACrefauthors}%
\unskip\
\newblock
\APACrefYearMonthDay{2004}{}{}.
\newblock
{\BBOQ}\APACrefatitle {The multiple-record systems estimator when registrations refer to different but overlapping populations} {The multiple-record systems estimator when registrations refer to different but overlapping populations}.{\BBCQ}
\newblock
\APACjournalVolNumPages{Statistics in medicine}{23}{}{2267--81}.
\newblock
\begin{APACrefURL} \url{https://doi.org/10.1002/sim.1818} \end{APACrefURL}
\newblock
\begin{APACrefDOI} \doi{10.1002/sim.1818} \end{APACrefDOI}
\PrintBackRefs{\CurrentBib}

\end{thebibliography}

\end{document}